\documentclass[a4paper,11pt]{article}
\pdfoutput=1 

\usepackage{jinstpub} 
\usepackage{subfig}

\title{\boldmath Deep learning for track recognition in pixel and strip-based particle detectors}

\author[1]{O.~Bakina}
\author[1]{D.~Baranov}
\author[1]{I.~Denisenko}
\author[2,1]{P.~Goncharov}
\author[1]{A.~Nechaevskiy}
\author[1]{Yu.~Nefedov}
\author[3]{A.~Nikolskaya}
\author[1]{G.~Ososkov}
\author[2]{D.~Rusov}
\author[3]{E.~Shchavelev}
\author[4,5]{S.~S.~Sun}
\author[4,5]{L.~L.~Wang}
\author[4]{Y.~Zhang}
\author[1]{A.~Zhemchugov}

\affiliation[1]{Joint Institute for Nuclear Research, \ protect \\Russia, 141980, Dubna, Moscow region, 6 Joliot-Curie St.}

\affiliation[2]{Dubna State University,\protect\\%
19 Universitetskaya St., Dubna, Moscow Region, 141982, Russia}

\affiliation[3]{St Petersburg State University,\protect\\%
7-9 Universitetskaya Emb., St Petersburg 199034, Russia}

\affiliation[4]{Institute of High Energy Physics CAS,\protect\\
19B Yuquan Road, Shijingshan District, Beijing, 100049, China}

\affiliation[5]{University of Chinese Academy of Sciences,\protect\\
19A Yuquan Road, Shijingshan District, Beijing, 100049, China}

\emailAdd{zhemchugov@jinr.ru}

\abstract{%
 The reconstruction of charged particle trajectories in tracking detectors is a key problem in the analysis of experimental data for high-energy and nuclear physics. The amount of data in modern experiments is so large that classical tracking methods, such as the Kalman filter, cannot process them fast enough. To solve this problem, we developed two neural network algorithms based on deep learning architectures for track recognition in pixel and strip-based particle detectors. These are TrackNETv3 for local (track by track) and RDGraphNet for global (all tracks in an event) tracking. These algorithms were tested using the GEM tracker of the BM@N experiment at JINR (Dubna) and the cylindrical GEM inner tracker of the BESIII experiment at IHEP CAS (Beijing). The RDGraphNet model, based on a reverse directed graph, showed encouraging results: 95\% recall and 74\% precision for track finding. The TrackNETv3 model demonstrated a recall value of 95\%  and 76\% precision. This result can be improved after further model optimization.}
  
\keywords{Track reconstruction, GEM detectors, Deep learning, Convolutional neural networks, Graph neural networks}

\arxivnumber{2210.00599} 

\begin{document}
\maketitle
\flushbottom

\section{Introduction}
\label{sec_1}
In physics experiments, tracking algorithms have evolved along with the development of experimental facilities and technologies for particle detection. Currently, the most effective tracking methods are based on the Kalman filter~\cite{Fruhwirth:1987fm}. A typical scheme of tracking using the Kalman filter is to find a track seed, then extrapolate it to the next coordinate detector and find a hit belonging to the track near the extrapolated point. Further, the procedure repeats, taking into account the newfound hit. This method naturally accounts for the inhomogeneity of the magnetic field, multiple scattering, and energy losses when a particle passes through the matter. However, these are the computational complexity of the Kalman filter and the significant difficulty of its parallel implementation that limits its data processing speed, especially when the track multiplicity is large. It is the reason that leads to the search for alternative methods for finding tracks to take advantage of modern hardware.

The first attempt to use artificial neural networks for track reconstruction was made by B.~Denby~\cite{Denby:1987rk}, using multilayer perceptrons and Hopfield's fully connected neural networks~\cite{Hopfield, HopfieldTank}. Later, the imperfection of these neural network algorithms, a sharp drop in tracking efficiency with an increase in the noise level, and the multiplicity of events led to the emergence of a modification of Hopfield's neural network algorithms, called elastic neural networks~\cite{Gyulassy:1990et}. Unfortunately, the efficiency of elastic tracking methods strongly depended on the proper choice of an initial approximation. Recently, deep learning methods have become a major focus of attention due to their ability to reveal hidden nonlinear dependencies in data and to the existence of an efficient parallel implementation of linear algebra operations lying in the base of these methods. The application of these methods for track reconstruction looks very promising, although algorithms for `deep tracking' that surpass classic methods, such as the Kalman filter, in performance have yet to be developed.

The TrackNETv3 and RDGraphNet track recognition algorithms, based on deep learning methods are proposed in this study to find tracks in pixel and strip particle detectors when a track should be recognized from a set of spatial point hits. The performance of these algorithms was studied in application to track finding in the GEM-based tracker of the BM@N experiment~\cite{Baranov:2017dtj} at JINR and in the CGEM-based inner tracker of the BESIII experiment~\cite{CGEM1} at IHEP CAS, Beijing.

\section{TrackNETv3 neural network}
\label{sec_2}
\begin{figure}[!ht]
  \begin{center}
    \includegraphics[scale=0.5]{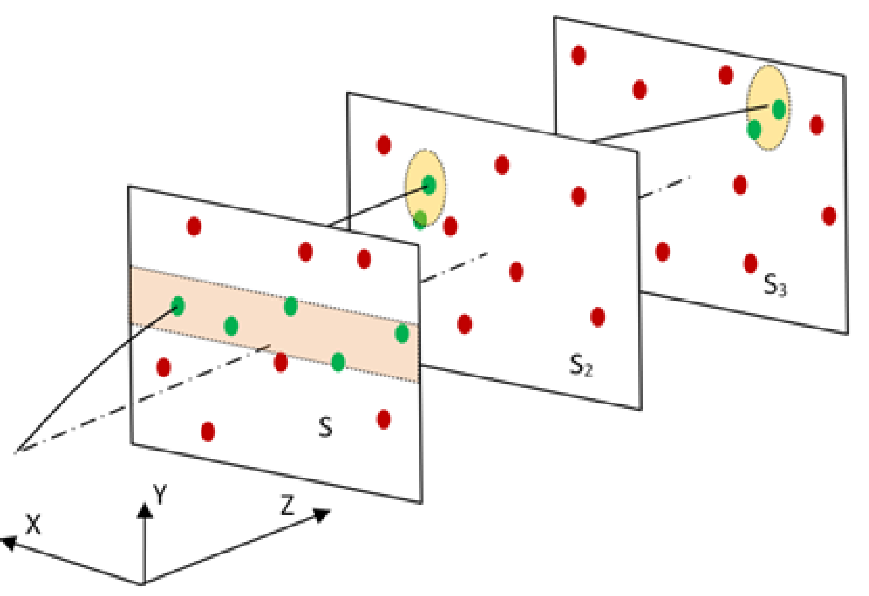}
    \caption{%
      Procedure of track finding in a fixed target experiment using the TrackNET algorithm.
    }
    \label{fig_2stage}
  \end{center}
\end{figure}

It is convenient to represent a typical tracking detector by a set of detector layers, cylindrical for a collider experiment or planes for a fixed target one (Fig.~\ref{fig_2stage}). A particle passing through detector layers produces hits in space. The layer position sets the longitudinal coordinate $Z$ of the hit, and the transversal position of the hit in the $X-Y$ plane is directly measured in pixel-based detectors or reconstructed from two or more stereo layers of strip-based detectors\footnote{When used in a collider experiment with coaxial tracker layers, cylindrical hit coordinates can be used instead: $Z = R, X = \phi, Y = z$, where $\phi$ is in the range of $[-\pi,\pi]$.}. Some of these hits can be noise or fake ones, which are inevitable in the case of strip-based readout due to the appearance of spurious strip intersections if the number of tracks is more than one. Having performed a combinatorial spatial search for track candidates simultaneously using two coordinate projections, one can use then a recurrent neural network (RNN) to separate true and false (ghost) tracks. This idea underlies the TrackNET group of algorithms, including TrackNETv1~\cite{Baranov2019}, TrackNETv2~\cite{Goncharov:2019moh}, and TrackNETv3, where each following algorithm is an evolution of the previous one. Since the first two algorithms have been described elsewhere, we only briefly summarize their properties below. The TrackNETv3 algorithm is new, so we discuss it further in more detail.

The two-stage tracking was successfully implemented using Gated Recurrent Units (GRUs) with a somewhat simplified gate mechanism for RNNs~\cite{Baranov2017}.  Both stages were combined into a single recurrent neural network, TrackNETv1,  to speed up the search of track candidates. Four more neurons were added to the classifier network, performing prediction of the elliptical region in the next layer, where it is necessary to look for the continuation of a track candidate. Two neurons predict the center of ellipse, and the other two do the same for the size of its semi-axes. One more, the fifth output neuron, classifies track candidate as a true track or a ghost one. The TrackNETv1 loss function is given by the equation:
\begin{gather}
  \label{eq_tracknet_v1_loss}
  J = \max{(\lambda_1, 1-p)}FL(p, p') + p\left( \lambda_2 \cdot \sqrt{\left(\frac{x-x'}{R_1}\right)^2 + \left(\frac{y-y'}{R_2}\right)^2}
    + \lambda_3 \cdot R_1 R_2 \right),
\end{gather}
where $\lambda_{1-3}$ are the weights; $p$ indicates whether the set of input hits belongs to a true track; $p'$ is the predicted probability of track/ghost; $x'$, $y'$ are the predicted coordinates of the ellipse center; $x$, $y$ are coordinates of the next hit of the true track segment; $R1$, $R2$ are semiaxes of the ellipse; $FL(p, p')$ is a balanced focal loss \cite{focalloss} with a weighting factor $\alpha \in [0, 1]$.

Combining track extrapolation with testing the hypothesis that the set of hits belongs to a true track and is compatible with a smooth track curve is a kind of multitask learning. Such learning strategy requires a careful choice of hyperparameters for the objective for minimization. The reason is that simultaneous optimization of multiple objectives makes the surface of the loss function more complex, causing problems in stability and convergence. At the same time, the regression part with ellipse prediction may work as a regularizer for the classification part preventing the network from rolling into local optimum by imposing the need to learn the spatial structure of tracks. Note, that this essentially reproduces the idea of the Kalman filter with the difference that it is a neural network that approximates the track parameters using synaptic weights determined during its training.

The TrackNETv1 model has some shortcomings, though. First, the TrackNETv1 loss function is too sensitive to its three parameters. Also, it turns out that the neural network is difficult to train in the case of strip-based readout with many fake hits. The problem is to find a balance between good extrapolation and classification when the number of false track candidates is substantially larger than the number of true ones. Keeping in mind, that predicting an ellipse to find the next hit of a track candidate already contains an information about the smoothness of the track curve, we decided to remove the classification from the loss function, coming to the TrackNETv2 model.

\begin{figure}[!ht]
  \begin{center}
    \includegraphics[scale=0.5]{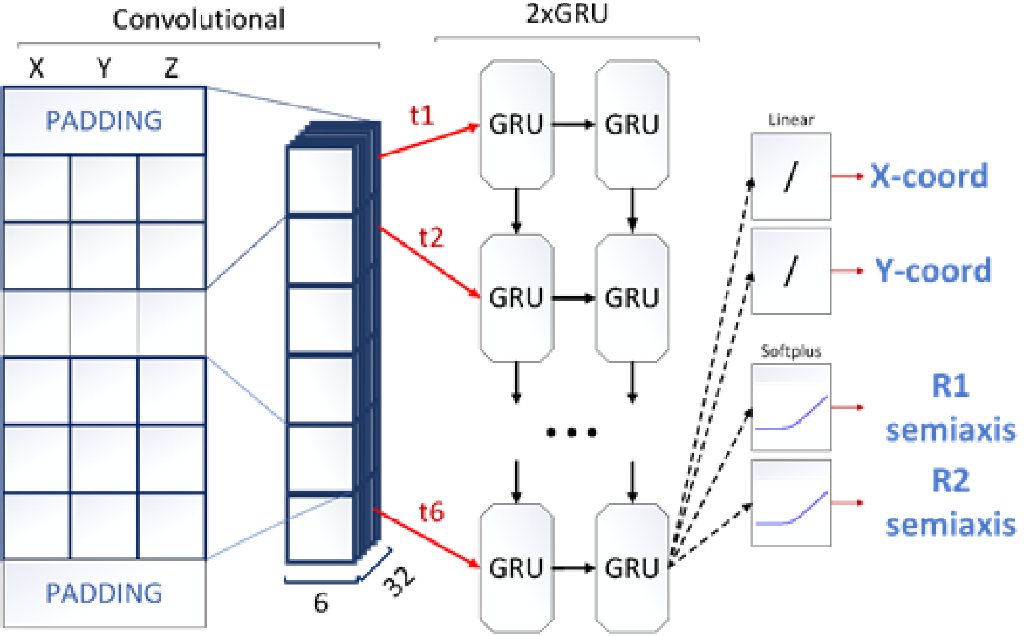}
    \smallskip
    \caption{Architecture of the TrackNETv2 model.}
    \label{fig_tracknet}
  \end{center}
\end{figure}

The architecture of the TrackNETv2 model is shown in Fig.~\ref{fig_tracknet}. The input of the neural network is a matrix, where each row contains hit coordinates for a track candidate. The first row of the matrix contains hit parameters in the first detector plane, the second row in the second one, and so on. The size of the matrix is fixed, and the number of rows corresponds to the number of detector planes, except one. The remaining input is padded by zeros. The loss function is described by the equation:
\begin{gather}
  \label{eq_1}
  J = \lambda_1 \cdot \sqrt{\left(\frac{x-x'}{R_1}\right)^2 + \left(\frac{y-y'}{R_2}\right)^2}
    + \lambda_2 \cdot R_1 R_2,
\end{gather}
where  $\lambda_1,\lambda_2$ are the weights; $x',y'$ is the predicted ellipse center; $x,y$ are the coordinates of the next hit of the current track; $R_1,R_2$ are the semi-axes of the predicted ellipse. We chose the linear activation function to predict the position of the ellipse center, while for semi-axes the softplus~\cite{softplus_act} activation function is used. Note, that Eq.~\ref{eq_1} is just a regression part of Eq.~\ref{eq_tracknet_v1_loss}. The loss function allows to predict the ellipse center close to the expected position of the next hit, simultaneously reducing the size of the ellipse to the extent the true hit is still inside. Thus, the track candidate gets longer until it reaches the last detector layer, or the next hit is missing in the predicted ellipse. 

Currently, the hyperparameters $\lambda_1$ and $\lambda_2$ should be tuned for every experiment independently, according to the input data, e.g. the reference system (polar vs cartesian) or normalization of coordinates (raw vs normalized), and the configuration of the tracking detector. Good starting point is $\lambda_1=0.8$ and $\lambda_2=0.2$.

Tracks can be of different lengths and contain different number of hits. This must be taken into account when training the model. One can use a special bucketing procedure, when tracks were sorted by length. However, such a procedure is unnatural for training RNNs, and requires the complicated bucket balancing. Instead, we do the following. First, the whole true track from simulated data, except the last hit, is passed to the input of the model. At the same time, all hits, except the first one, are used as labels for loss function calculation. The model makes predictions for each hit, e.g. for the first hit it will try to predict the ellipse where to search for the second hit, for the first two hits the model tries to guess where the third hit its, etc. Thereby the network predicts ellipses for each timestamp, and we calculate the loss for each model prediction. Then we average the loss value across all timestamps and perform backpropagation. Such a strategy of training is called many-to-many, when the model generates output for every hit in the input sequence.

There can be samples of different sizes in a single batch of inputs, and we need to pad shorter track candidates to the length of the longest ones across the batch. Usually the padding value equals zero. For instance, if the batch includes two samples with length 4 and length 6, then the candidate with four hits will be extended by another the two hits with all three coordinates equal zero. During loss estimation, a special mask for padding is calculated, and these timestamps are excluded from the optimization process.

Switching the training strategy of the TrackNETv2 model forced us to revise the original architecture by removing the convolutional layer. For using convolution, the `same' padding strategy is necessary for the correct number of predictions. For a convolution kernel size of three, the padding value is one. The standard way of applying padding, namely, one zero value on each side, is incorrect, because it results in the violation of the order of predictions: at the first prediction, the model will use the second time hit, and so on. Thus, if we try to train the network using the many-to-many strategy, the standard convolution will promote the model to cheat during training, and  the evaluation will fail. To solve this problem, we could make the padding operation causal - add the necessary amount of zero elements (two, in our case), but only at the beginning. However, our study showed that such causality performs worse than a simple drop of the convolutional layer.

To speed up the TrackNETv3 inference, we use Faiss library~\cite{faiss} and make a search index of hits. This index provides optimal finding of $K$ hits nearest to the ellipse center, where $K$ is a parameter of the model. We estimate the complexity as $O(NK)$, where $N$ is the total number of hits. This estimate is valid if the condition $K << N$ is fulfilled.  

All these improvements have become a further development of the network to version TrackNETv3.

The TrackNETv3 network usually demonstrates high ($\sim$~99\%) recall, but precision, i.e. the number of ghost tracks, strongly depends on the conditions of the experiment. An additional classifier can be added to the network, to counter this problem and improve precision. An example of the classifier like this is described in section~\ref{section_bes3}, in relation to the BESIII-CGEM detector.

Finally, we sum the properties the TrackNET group of algorithms up in the Table~\ref{tab_tracknet_versions}.

\begin{table}[!ht]
  \begin{center}
    \caption{%
      Properties of algorithms of the TrackNET family.
    }
    \label{tab_tracknet_versions}
    \begin{tabular}{|p{0.15\textwidth}|p{0.15\textwidth}|p{0.15\textwidth}|l|p{0.1\textwidth}|c|}
      \hline
TrackNET version & Learning problem & Training strategy & Layer types & $k-NN$ hit search & Track classifier \\ \hline
TrackNETv1 & classification + regression & many-to-one & CNN + RNN & No & No \\ \hline
TrackNETv2 & regression & many-to-one & CNN + RNN & No & No \\ \hline
TrackNETv3 & regression & many-to-many & RNN & Yes & Yes \\ \hline
    \end{tabular} 
  \end{center}
\end{table}

\section{RDGraphNet neural network}
Being a local tracking method, the TrackNETv3 model does not allow one to assess the global picture of an event, to see the dependence between individual tracks or groups of tracks and directly recognize phenomena such as secondary vertices. One may also expect the linear growth of tracking time with the increase of the track multiplicity in the event. To overcome these difficulties, a global tracking method based on a Graph Neural Network (GNN) was developed~\cite{Shchavelev}.

\begin{figure}[!ht]
  \begin{center}
    \includegraphics[scale=0.5]{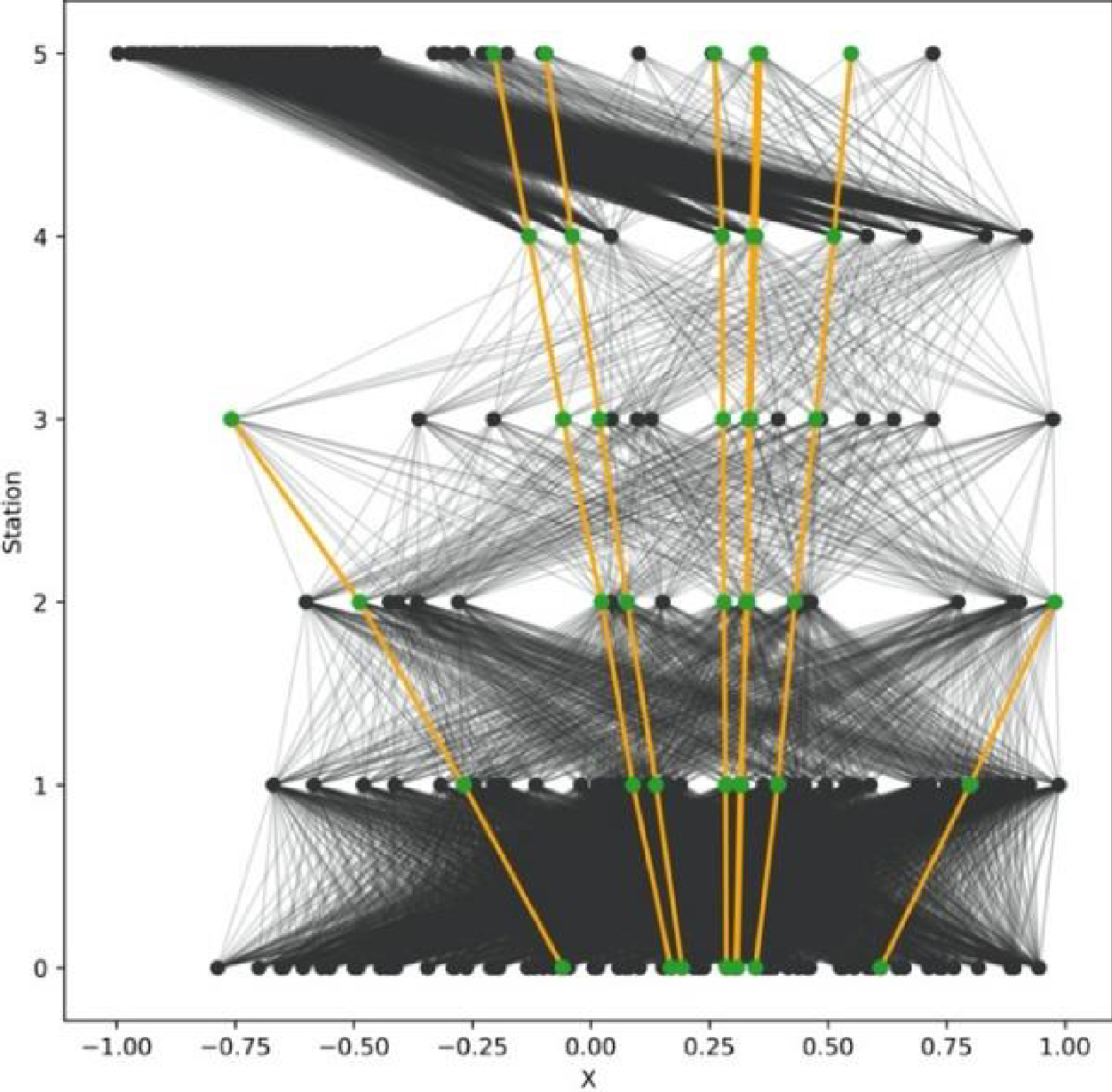}
    \smallskip
    \caption{%
        Event in the BM@N detector, represented as a graph. The green nodes are true hits, the black nodes are fake hits produced by strip crossings in the GEM-based detector. The orange edges represent real track segments, all the other edges are shown in black.}
    \label{fig_graph_bmon}
  \end{center}
\end{figure}

An event represents a directed graph. Hits play the role of graph nodes. All hits located at adjacent detector layers are connected by edges (track segments).  The nodes within one detector layer are not connected. As an illustration, an event in the BM@N detector, represented as a directed graph is shown in Fig.~\ref{fig_graph_bmon}.

The graph can be expressed in the form of four matrices: 
\begin{itemize}
  \item
    $X$ is the matrix of the parameters of nodes of size $N \times M$,
    where $N$ is the number of nodes, and $M$ is the number of parameters.
    The hit coordinates are used as node parameters, i.e. $M = 3$.
  \item
    $R_{\rm i}$ is the matrix of incoming edges of size $N \times E$ ($E$ is the number of edges).
    In this matrix $R_{\rm i}[i,j] = 1$ if the edge with the index $j$ is included in the node
    with the index $i$, otherwise it equals to zero.
  \item
    $R_{\rm o}$ is the matrix of outgoing edges, coming from the corresponding nodes. The matrix has
    the same properties as $R_{\rm i}$.
  \item
    $Y$ is the vector with the size $E$. $Y[j] = 1$ if the edge with the index $j$ 
    belongs to the real track, and zero otherwise. 
\end{itemize}

\begin{figure}[!ht]
  \begin{center}
    \includegraphics[scale=0.6]{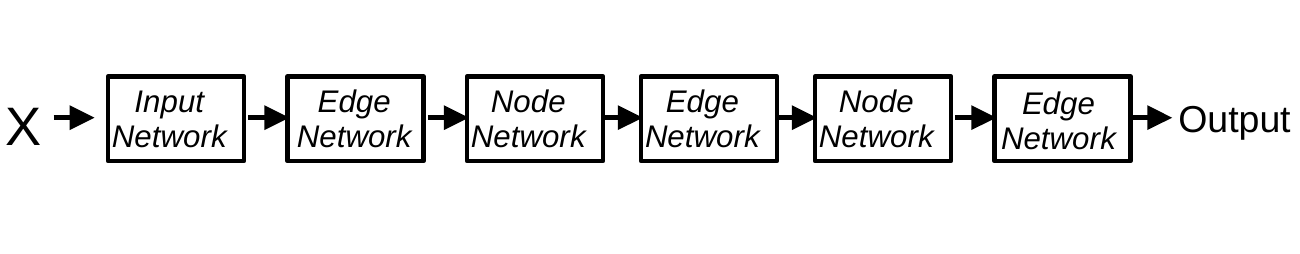}
    \smallskip
    \caption{%
      The scheme of GNN. The input matrix comes to the Input Network and then goes through the chain of two Edge-Node iterations. 
    }
    \label{fig_schemagnn}
  \end{center}
\end{figure}

There are three main components in the GNN: the Input, Node and Edge networks. The Input network is a multilayer perceptron (MLP) with a hyperbolic tangent activation function. The $X$ matrix is fed to the Input network. The goal of the network is to extract from the position of hits some features that will be used later in Edge and Node networks. The output of the Input network is fed to several iterations of `Edge-Node' networks. Any Edge network is a two-layer fully connected neural network whose task is to calculate the weights of the edges of the graph based on the features of the nodes associated with the corresponding edge. The activation function in the Edge network is a hyperbolic tangent, and the output layer uses a sigmoid activation function to determine whether a graph edge is a true track segment. The Node network is constructed in a similar way, but has a different purpose. This network recalculates the features of the nodes of the graph using the features of neighboring nodes and the weights of all edges coming to and going out from the node, which are obtained by the previous Edge network (see Fig.~\ref{fig_schemagnn}). Finally, the Edge network defines whether graph edges correspond to true tracks or ghosts, and works as the output layer of the GNN.

Thus, each `Edge-Node' iteration propagates the information of every node along the edges and combines it with the content of other nodes. The number of such iterations is a hyperparameter of this model. A large number of iterations increases the amount of computation and reduces the network performance, but at the same time, leads to better network convergence.

The authors of~\cite{Goncharov2020} have shown that a straightforward adaptation of the GNN model developed for experiments on the LHC pixel detectors~\cite{cerngnn}, does not allow to achieve the acceptable accuracy for a strip-based detectors due to the presence of a large number of fake hits. However, it is possible to come up with a way to preprocess the event so as to significantly reduce the number of false track segments in the graph before feeding it to the GNN. A preprocessing algorithm of building a reverse directed graph (reverse digraph) has been introduced in~\cite{Shchavelev}. The edges of a new graph $G_r$ are the nodes of the original graph $G$, and the nodes of the reverse graph are the edges of the original one. Effectively, track segments in the reverse digraph connect three hits instead of two in the original graph, so the GNN receives more information from each node of the input graph and learns faster.

As a result, this approach reduces the number of fake track segments by an order of magnitude, while preserving all true track segments. To achieve this, the weight for each edge of the graph $G_r$ is calculated by the formula: 
\begin{gather}
  \label{eq_5}
     w_i = \sqrt{ (dZ_{i+1}-dZ_i)^2 +
                  (dX_{i+1}-dX_i)^2 +
                  (dY_{i+1}-dY_i)^2},
\end{gather}
where $dZ_j = Z_j-Z_{j-1}$,  $dX_j = X_j-X_{j-1}$, $dY_j = Y_j-Y_{j-1}$ for $i=1$, $j \in \{1,2\}$ and $Z_j$, $X_j$ and $Y_j$ are the hit coordinates. One can notice that an edge of the $G_r$ now contains the information of the 3 consecutive hits of the initial graph $G$. During training, at the input, the network receives a reverse digraph along with labels indicating whether the corresponding edge is true (connects three true hits) or fake (contains one or more fake hits). The training dataset contains only those edges whose weight is $w < 0.073$. This value was chosen during the statistical analysis of training data to keep the maximum number of true edges while minimizing the number of fake edges of the $G_r$ (by the construction of the algorithm an edge of $G_r$ is true if it contains 3 true hits of the $G$). After training, the RDGraphNet (Reversed Directed Graph Neural Network) neural network predicts the value $x \in [0,1]$. The true edges of the track candidate are those edges for which $x$ exceeds a certain threshold. Again, the threshold is used for the adjustment of the efficiency and purity of track recognition. In this work, the threshold is fixed and equal to $0.5$. The RDGraphNet hyperparameters used for track finding in the planned CGEM detector of the BESIII experiment are summarized in Table~\ref{tab_pargraphnet}.
\begin{table}[!ht]
  \begin{center}
    \caption{Training hyperparameters of RDGraphNet.}
    \label{tab_pargraphnet}
    \begin{tabular}{|p{0.5\textwidth}|p{0.35\textwidth}|}
      \hline
Parameter & Value \\ \hline
Number of Node-Edge iterations & 2 \\ \hline
Dimension of input features & 5  (coordinates
      $r_j,r_{j-1}$, $\varphi_j,\varphi_{j-1}$
      \& station serial number ) \\ \hline
Number of neurons on the hidden MLP layer & 96 \\ \hline
    \end{tabular}
  \end{center}
\end{table}

\section{Implementation of neural networks and Ariadne library}
\begin{figure}[!ht]
  \begin{center}
    \smallskip
    \includegraphics[scale=0.6]{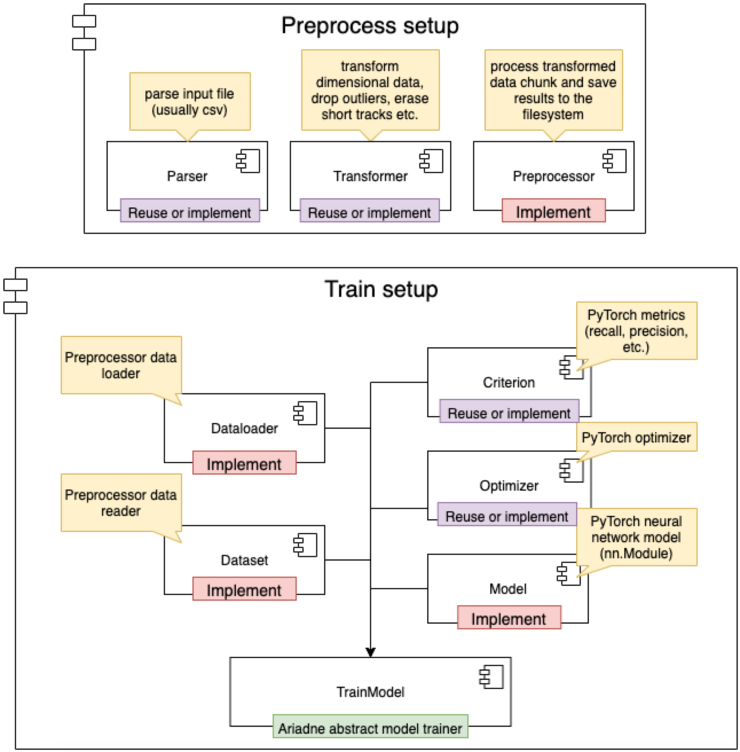}
    \smallskip
    \caption{%
      General Ariadne setup process.
    }
    \label{fig_xxx}
  \end{center}
\end{figure}

The design and performance study of `deep tracking' algorithms described in this work was greatly simplified by using the Ariadne toolkit~\cite{ariadne_git} developed by the authors. The Ariadne toolkit is an open-source Python3 library that aims at solving complex high-energy physics tracking tasks with the help of deep learning methods. Ariadne provides the researcher with a handy framework for the rapid prototyping of a new neural network model for event reconstruction tasks and a step-by-step fully reproducible pipeline including data preparation, training and evaluation phases. Ariadne can use modern deep learning libraries (such as a PyTorch~\cite{pytorch_lib}, Pytorch Lightning~\cite{pytorch_lightning_lib}) and allows utilizing multi-core CPU and distributed multi-GPU facilities during preprocessing, training, and performance evaluation. The general structure of the toolkit is shown in Fig.~\ref{fig_xxx}.

The toolkit consists of several modules that allow one to extend it with the details of any experiment setup, to define a neural network model, and describe the user's data structure, so that one can train and then study the performance of the model and compare it with other models being part of Ariadne. The neural networks described in this work are already implemented in the Ariadne toolkit.

\section{Performance study}

\subsection{Tracking detector of the BM@N experiment}


BM@N is a fixed target experiment at the Nuclotron at JINR, aimed at studying the properties of baryon matter formed by the collision of heavy ions at beam energies from 2 to 6 GeV~\cite{bmn_run7}. For this study, the RUN 6 and RUN 7 configurations of the BM@N setup were used. RUN 6 configuration of the setup contains a central tracker comprising 6 detector planes (stations) of GEM-based strip detectors. Each station has 2 stereo layers with straight and inclined (5-15 degrees) strips. Carbon beam interaction was simulated in the BM@N detector using the BmnRoot framework~\cite{batyuk2019bmnroot} and the LAQGSM event generator. In total, 550 thousand carbon-carbon (C + C) interactions at a beam energy of 4 GeV were generated. 150 thousand events were considered as a test sample, and the rest of the events were used for training.

Two metrics are used to characterize the performance of the algorithms:

\begin{itemize}
    \item Hit efficiency: share of true hits found by a network out of all true hits in a single event.
    \item Track efficiency: share of full tracks without gaps found by a network out of all tracks in a single event.
\end{itemize}

The results are shown in Table~\ref{tab_results_bmn_run6}.

\begin{table}[!ht]
  \begin{center}
    \caption{%
      Model evaluation results.
    }
    \label{tab_results_bmn_run6}
    \begin{tabular}{|p{0.3\textwidth}|c|c|}
      \hline
\ & RDGraphNet & TrackNETv2 \\ \hline
Track efficiency (\%) & ~87 & 95.93 \\ \hline
Hit efficiency (\%) & ~94 & 99.18 \\ \hline
    \end{tabular}
  \end{center}
\end{table}

RUN 7 has a configuration of the detector with 6 GEM stations and 3 silicon stations before the GEM ones, 9 detector planes in total. We generated about 750K events for training and validating the model. We used events with a beam energy equal to 3.2 GeV. Interactions with an argon beam and a plumbum target (ArPb) were considered. The magnet amperage was set to 1250 A. The target was generated with the following parameters: X (mean: 0.7 cm, std: 0.33 cm), Y (mean: -3.7 cm, std: 0.33 cm), Z (center: -1.1 cm, thickness: 0.25 cm). The multiplicity of events varied up to 100 tracks per event, and 37 tracks per event on the average. The number of hits can be more than 500 per station, and about 68\% of all hits were fakes. 

At the moment, we have trained on BM@N RUN 7 data only the first part of the TrackNETv3 model without a classifier. We normalized the hit coordinates in the dataset depending on the boundaries of the detector area and balanced the size of groups with tracks of different length. In additional, we changed the value of the alpha parameter of the loss function which weights the cost of ellipse fitting to the loss function in opposition to the ellipse area. The optimal value for the “normalized” dataset should be much smaller, otherwise the predicted ellipses will be too large. We set alpha at 0.01.

Two metrics were used to measure the quality of the model:
\begin{itemize}
  \item
    Recall: $N_\text{true}^\text{rec}/N_\text{MC}$, where
    $N_\text{true}^\text{rec}$ is the number of correctly reconstructed true tracks and
    $N_\text{MC}$ -- total number of all modelled real tracks (Monte-Carlo).

  \item
    Precision: $N_\text{true}^\text{rec}/N^\text{rec}$, where
    $N^\text{rec}$ is the total number of reconstructed tracks (including fake tracks).
\end{itemize}
From the point of view of searching for tracks, the concept of recall coincides with the concept of track recognition efficiency, and precision  coincides with the purity of track recognition. The probability that the track found is fake is given by the expression $1-p$, where $p$ is the precision. The track in our experiment is considered to be correctly reconstructed if 70\% of hits or more are found as in the corresponding Monte-Carlo track. 

The best results were achieved using 5 nearest neighbors as candidates for entrance in the predicted ellipses. The results are shown in Table~\ref{tab_results_bmn_run7}. Time measurements were performed using the Intel (R) Xeon (R) Gold 6148 CPU @ 2.40~GHz for a version of a single-threaded program implemented in Python. Such a large time spread is explained by the fact that in some events there are several hundred thousand candidate tracks that the model should process. The processing can be made much faster by translating the program to a multi-threaded C++ realization.

\begin{table}[!ht]
  \begin{center}
    \caption{%
      TrackNETv3 model evaluation results on BM@N RUN 7 data.
    }
    \label{tab_results_bmn_run7}
    \begin{tabular}{|p{0.4\textwidth}|c|c|}
      \hline
Track Efficiency (recall) (\%) & 98.30 \\ \hline
Track Purity (precision) (\%) & 2.09 \\ \hline
Mean time to process one event (s) & 0.1004 $\pm$ 0.1299 \\ \hline
    \end{tabular}
  \end{center}
\end{table}

\subsection{Inner tracker of the BESIII experiment}
\label{section_bes3}

BESIII~\cite{Ablikim:2009aa} is an experiment running at the BEPCII  $e^+e^-$ collider at the Institute of High-Energy Physics in Beijing. The physics program of the BESIII experiment ~\cite{Asner:2008nq} covers a wide range of problems, including the study of the $\tau$-lepton, charmed particles and charmonium states. In the BESIII detector, tracks are detected in a drift chamber that consists of two parts: outer and inner one. Currently, the replacement of the inner part, based on Cylindrical GEM detectors, is being developed. The CGEM inner tracker consists of three layers of GEM detectors. The readout anode of each CGEM layer is segmented with 650~µm pitch $XV$ patterned strips. $X$-strips are parallel to the CGEM axis and provide the $\phi$ coordinate. $V$-strips, which have a stereo angle of 30 or 45 degrees to $X$-strips, define the $Z$ coordinate.

The presented methods were applied to reconstruct tracks in the CGEM-IT detector of the BESIII experiment at a collision energy of 3.686 ~ GeV. We utilized data from the Monte Carlo simulation of electron-positron annihilation events with the formation and decay of the $\psi (3686)$ resonance. For each event, the Monte Carlo simulation produces hits that are either associated with true charged particle tracks or marked as fakes. The procedure for modeling hits is described in detail in work ~\cite{Denisenko:2019vno}. We used events in which the modeled track always contains three hits, and the tracks do not overlap with each other. The training and validation dataset consisted of 250,000 simulated events. In the dataset, typically there are more than 3 fake hits per one true hit, which complicates the track recognition. Table~\ref{tab_dataset} contains dataset characteristics for RDGraphNet and TrackNETv3. Note, that only part of the dataset is used for training, and the other part formed the validation dataset. 

\begin{table}[!ht]
  \begin{center}
    \caption{%
      Characteristics of datasets used for training and validation. 
    }
    \label{tab_dataset}
    \begin{tabular}{|p{0.5\textwidth}||r|r|}
      \hline
Characteristics & RDGraphNet  & TrackNETv3 \\ \hline\hline
\multicolumn{3}{|c|}{Training}\\ \hline
Total no. events  & 215453 & 134997\\ \hline
Total no. tracks  & 1095237 & 686246 \\ \hline
No. hits (including real and fake ones) & 11391914 & 7137867 \\ \hline
Fraction of fake hits  & 77\% & 77\% \\ \hline
Fraction of real hits & 23\% & 23\% \\ \hline\hline
\multicolumn{3}{|c|}{Validation} \\ \hline
Total no. events & 4326  & 4326 \\ \hline
Total no. tracks  & 19,653 & 19,653 \\ \hline
No. hits (including real and fake ones) & 225321 & 225321 \\ \hline
Fraction of fake hits  & 77\% & 77\% \\ \hline
Fraction of real hits & 23\% & 23\% \\ \hline
    \end{tabular}
  \end{center}
\end{table}

The parameters used in training neural networks are presented in Table~\ref{tab_teaching}.

In the case of BESIII-CGEM, with few detector layers and a large number of fake hits, the number of ghost tracks becomes unacceptably high. The reason is the high probability to pick up a combination of few fake hits compatible with the track hypothesis. To solve this problem, information about the beam interaction point, giving the primary vertex of the track, would be helpful. However, the model can be trained to use the primary vertex, since this information is implicitly present in the data. To do this, TrackNETv3 was supplemented with another classifier, which was trained to give the probability of a track candidate to be a real track. 

\begin{figure}[!ht]
  \begin{center}
    \subfloat[]{\includegraphics[width=0.3\textwidth]{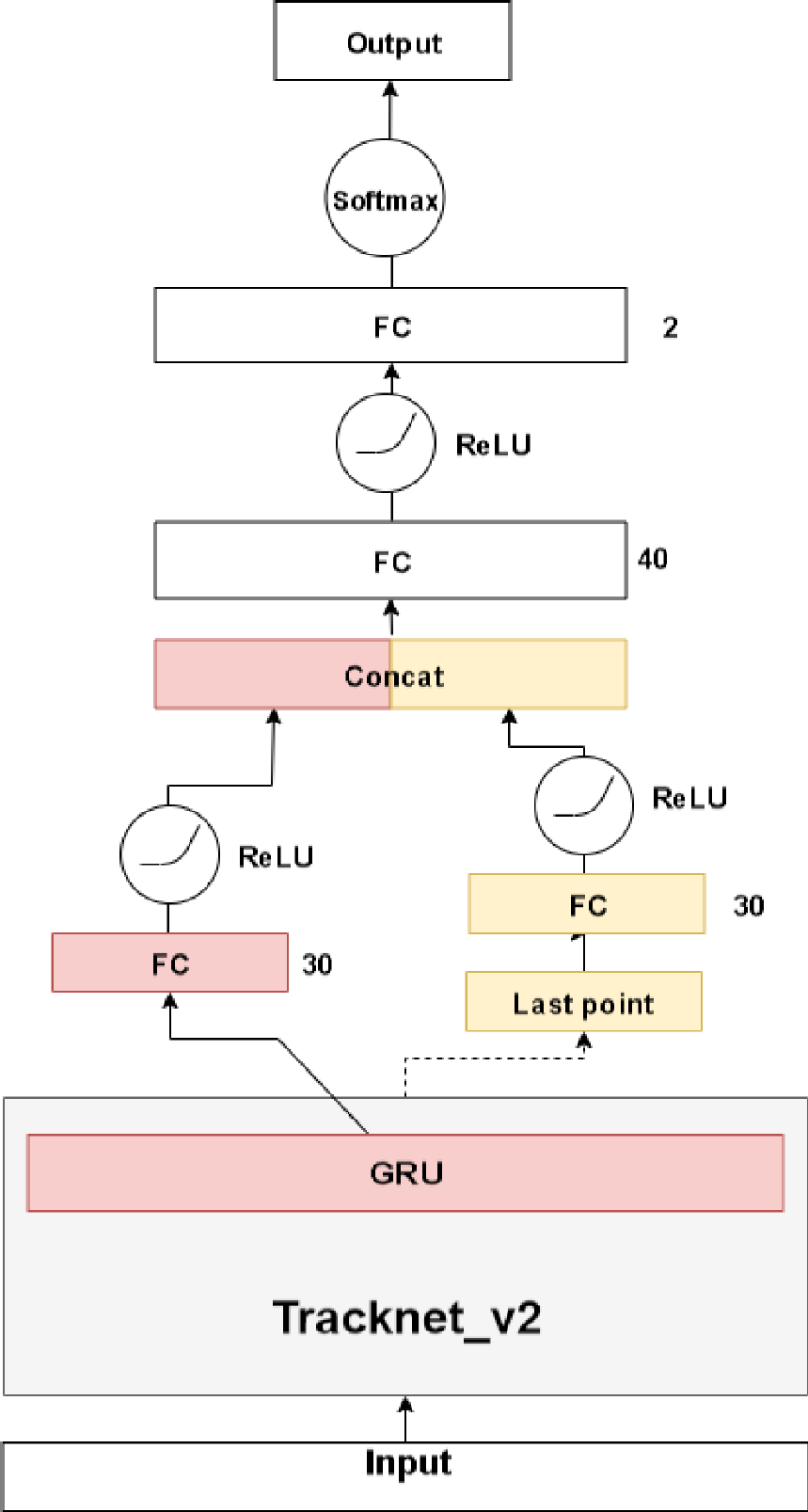}}
    \qquad
    \subfloat[]{\includegraphics[width=0.35\textwidth]{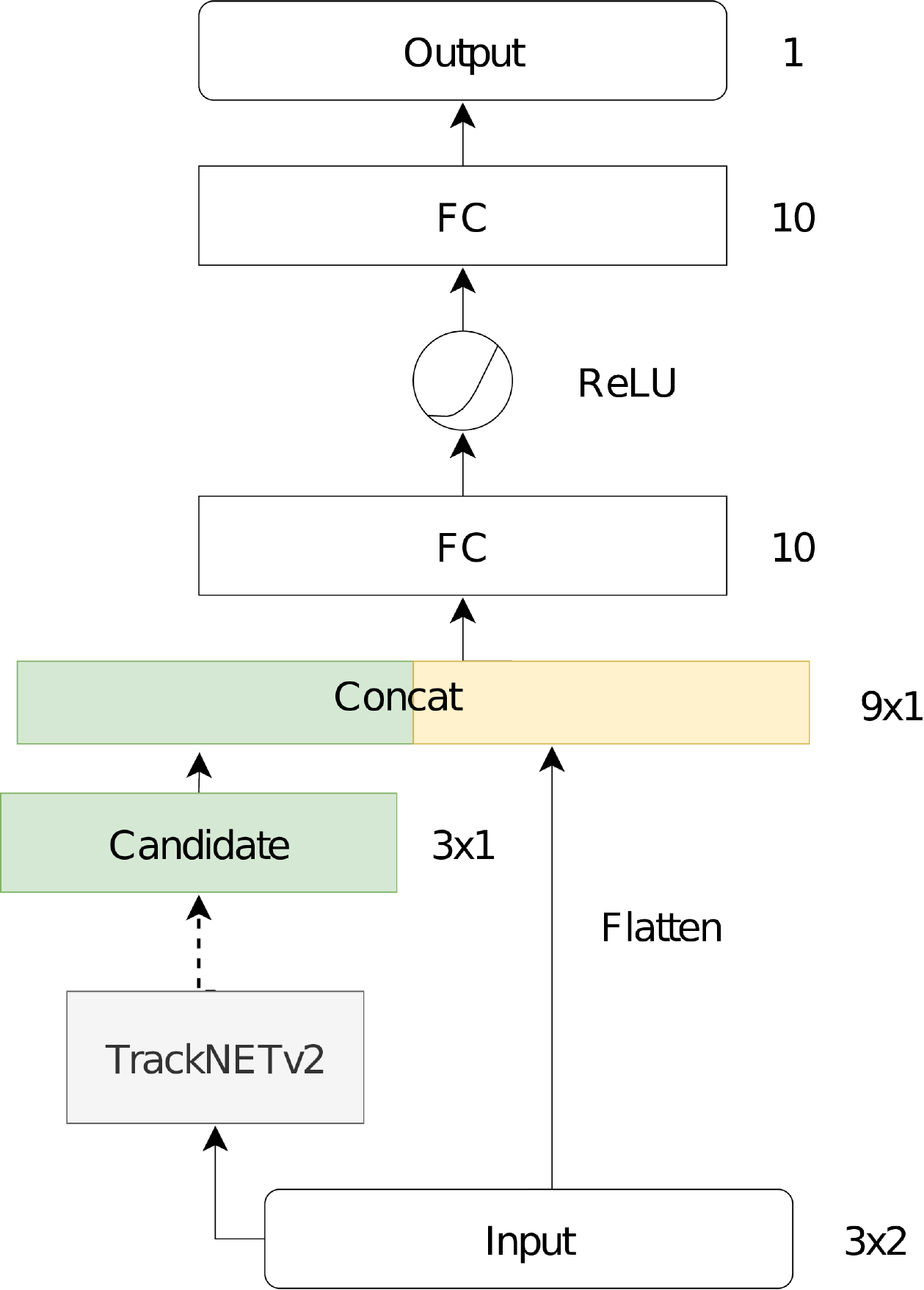}}
    \smallskip
    \caption{%
      Scheme of the Classifier-GRU (a) and Classifier-Coords (b) classifiers.
    }
    \label{fig_classifier}
  \end{center}
\end{figure}

Two different classifiers were tried (see Fig.~\ref{fig_classifier}). For the first classifier (Classifier-GRU), two inputs are used: the output of the second GRU layer of TrackNETv2 and the set of hit coordinates at the end of the track candidate (Last point). Each input is propagated to a fully connected layer (FC). Then the resulting vectors are concatenated (Concat) and transferred to another fully connected layer (FC). The ReLU activation function~\cite{relu_act} is used in all these network layers. The output of the last FC layer of the classifier uses the softmax~\cite{softmax_act} activation function.

The input of the second classifier (Classifier-Coords) contains the coordinates of three hits in a track candidate. The first two hits are input for the TrackNETv3 network and the third is the predicted one. These coordinates are concatenated into a vector of size 9 and then fed into a small fully connected block of two layers with ReLU activation. The output of the last FC layer uses the sigmoid ~\cite{sigmoid_act} activation function.

To train the classifiers, the proper choice of the loss function is crucial. The binary cross entropy ~\cite{crossentropy} is the standard for binary classification:
\begin{gather}
  \label{eq_3}
  BCE = - \sum{y_t  log(p_t)},
\end{gather}
where $y_t$ is the label of the t-th sample, and $p_t$ is the predicted probability. However, the use of this function gives the preference to easy cases with a high probability to be a true track and leads to a weak learning ability. The focal loss function solves this problem by adding the modulating factor $\gamma$ to the cross entropy value ~\cite{focalloss}. The idea is that if a sample is already well classified, we can significantly decrease the weight of its contribution to the loss. To set a balance of positive and negative examples, the weighting parameter $\alpha$ is used; it has the meaning of the inverse class frequency.
\begin{gather}
  \label{eq_4}
  FL = - \sum{ y_t  (1 - p_t)^\gamma  log(p_t)},
\end{gather}

TrackNETv3 with different combinations of classifiers and loss functions was tested and the results are shown in Table~\ref{tab_tracknet_clfs}. The results for the TrackNETv3 network without classifier, tested on the same set are also demonstrated. The need for an additional classifier becomes obvious. At the same time, one can see that different combinations of the classifier and the loss function give different results, although the recall is quite high for all. Nevertheless, the combination of Classifier-GRU and focal loss function is the best and we use it further.

\begin{figure}[!ht]
   \begin{center}
     \includegraphics[scale=0.45]{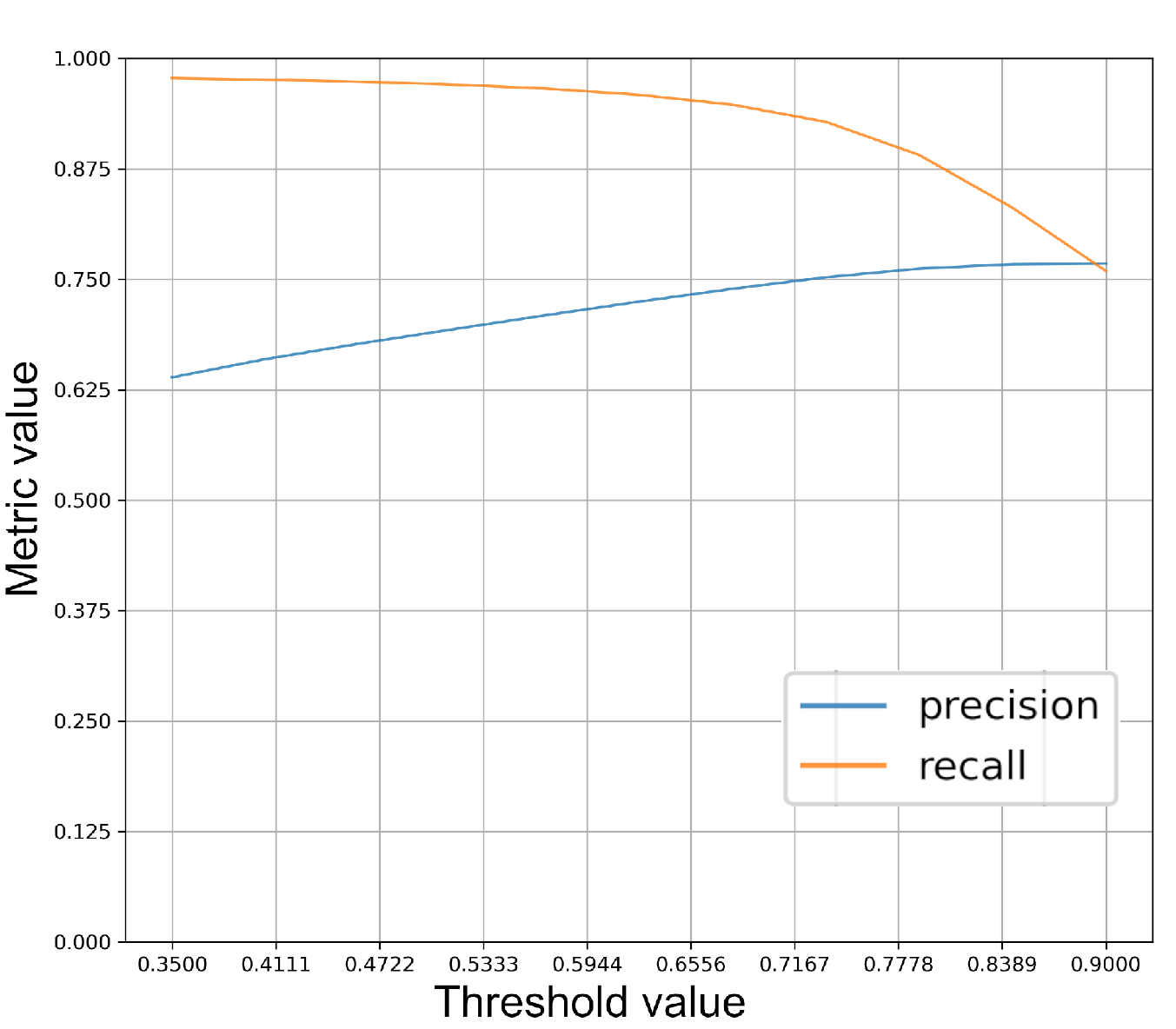}
     \smallskip
     \caption{%
     Dependence of the TrackNETv3 prediction quality on the threshold value, for Classifier-GRU and focal loss function.}
     \label{thresholds}
   \end{center}
\end{figure}

\begin{table}[!ht]
  \begin{center}
    \caption{%
      Comparison of TrackNETv3 performance with different classifiers and loss functions.
    }
    \label{tab_tracknet_clfs}
    \begin{tabular}{|p{0.3\textwidth}|p{0.2\textwidth}|p{0.15\textwidth}|p{0.15\textwidth}|}
      \hline
Classifier & Loss function & Precision & Recall \\ \hline
No classifier & TrackNETLoss & 0.01 & 0.99 \\ \hline
Classifier-GRU & Cross Entropy & 0.54 & 0.95 \\ \hline
Classifier-GRU & Focal & 0.677 & 0.98 \\ \hline
Classifier-Coords & Cross Entropy & 0.66 & 0.94 \\ \hline
Classifier-Coords & Focal  & 0.58 & 0.96 \\ \hline
    \end{tabular} 
  \end{center}
\end{table}
Since any classifier predicts the probability of a track candidate to be a real track, the proper threshold choice allows adjusting the efficiency and purity of track recognition (see Fig.~\ref{thresholds}).
In this study, we used the threshold of 0.7. Table~\ref{tab_results} shows the results of evaluating the trained models. 

\begin{table}[!ht]
  \begin{center}
    \caption{%
      Hyperparameters used for training.
    }
    \label{tab_teaching}
    \begin{tabular}{|p{0.18\textwidth}|c|c|}
      \hline
\ & RDGraphNet  & TrackNETv3 \\ \hline
optimizer & \multicolumn{2}{|c|}{Adam ~\cite{adamopt}} \\ \hline
loss function & \texttt{\small torch.nn.functional.binary\_cross\_entropy} & TrackNETv3 loss \\ \hline
learning rate & 0.0007 & 0.001 \\ \hline
fakes reweighting & $W_{\rm false} = w_{\rm false} \cdot 0.555$
                       & $W_{\rm false} = w_{\rm false} \cdot 0.625$ \\
real hits reweighting & $W_{\rm true} = w_{\rm true} \cdot 3$
               & $W_{\rm true} = w_{\rm true} \cdot 2.5$ \\ \hline
    \end{tabular}
  \end{center}
\end{table}

\begin{figure}[!ht]
  \begin{center}
    \smallskip
    \includegraphics[scale=0.4]{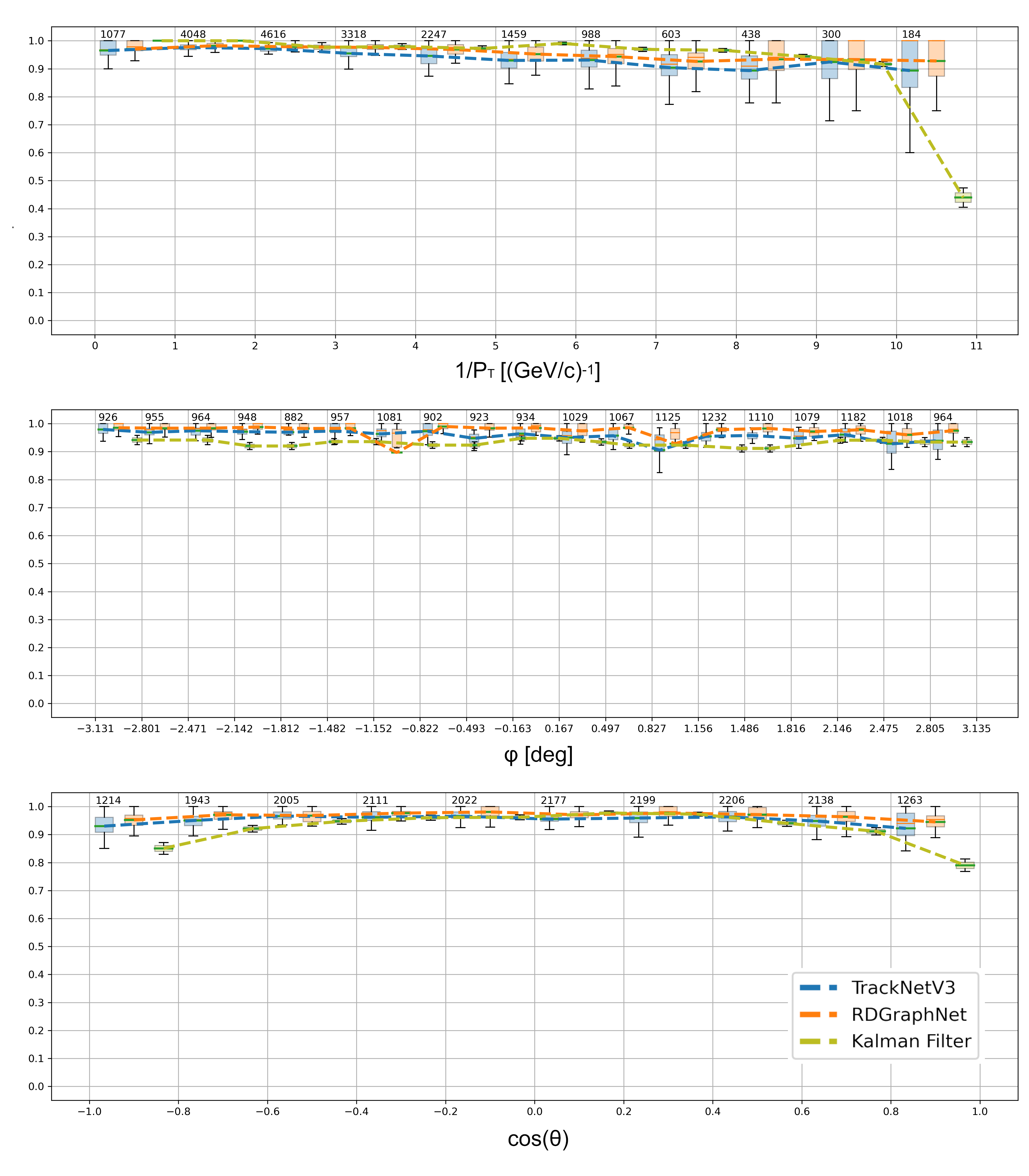}
    \smallskip
    \caption{%
      Track recognition efficiency vs inverse value of transverse momentum (top), azimuthal angle (middle) and cosine of the track polar angle (bottom). Total number of events equals to 4330. Results obtained using TrackNETv3 (blue line), RDGraphNet (red line) and the Kalman filter (green line) are shown.
    }
    \label{fig_efficiency}
  \end{center}
\end{figure}

\begin{figure}[!ht]
  \begin{center}
    \smallskip
    \includegraphics[scale=0.4]{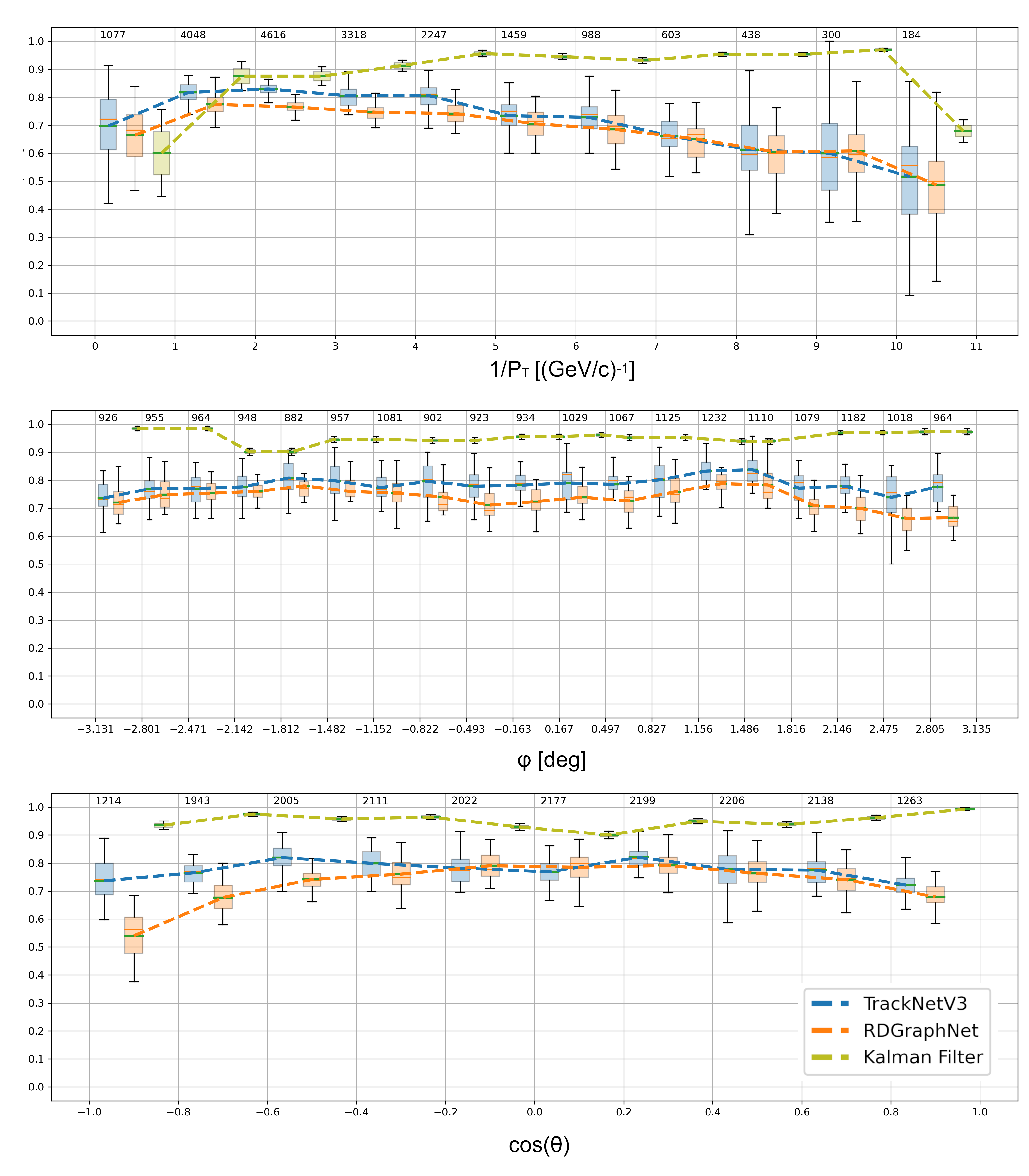}
    \smallskip
    \caption{%
      Track recognition purity vs inverse value of transverse momentum (top), azimuthal angle (middle) and cosine of the track polar angle (bottom). Total number of events equals to 4330. Results obtained using TrackNETv3 (blue line), RDGraphNet (red line) and the Kalman filter (green line) are shown. 
    }
    \label{fig_purity}
  \end{center}
\end{figure}

A track in our experiment always consists of three hits and is considered to be correctly reconstructed if all hits are found as in the corresponding Monte-Carlo track. Fig. \ref{fig_efficiency}-\ref{fig_purity} show the dependency of purity and efficiency on the inverse value of the transverse momentum, azimuthal angle and cosine of the track polar angle in comparison with the Kalman filter.

\begin{table}[!ht]
  \begin{center}
    \caption{%
      Model evaluation results.
    }
    \label{tab_results}
    \begin{tabular}{|p{0.3\textwidth}|c|c|}
      \hline
\ & RDGraphNet & TrackNETv3 \\ \hline
Track Efficiency (recall) & 0.9548 & 0.9475 \\ \hline
Track Purity (precision) & 0.7404 & 0.7594  \\ \hline
Events fully reconstructed & 0.8271 & 0.8081  \\ \hline
    \end{tabular}
  \end{center}
\end{table}

As for the track recognition efficiency, neural network methods do very well and are almost comparable to the classical algorithm. Remarkably, the neural network methods successfully solve the problem of tracks crossing the azimuthal boundary ($\phi=\pi=-\pi$), which breaks the continuity of about 1.5\% of the tracks. In addition, the neural network turns out to be more efficient for tracks close to the beam direction than the Kalman filter. This could be, of course, the result of optimizing the network between efficiency and purity. In this case, the purity for these polar angle regions should slightly decrease.

As for the track recognition purity, we clearly see that neural networks perform worse than the Kalman filter. We believe that one of the reasons may be a well elaborated deterministic track model embedded in the Kalman filter and contributing to more accurate extrapolation, while the neural network is forced to build this model by training and the accuracy of track extrapolation may turn out to be worse. Consequently, the probability of associating a false hit with a track increases. In this case, the performance of neural network methods can be greatly improved by properly training the network.

Still, these results are rather encouraging. First, `deep learning' tracking methods deserve further optimization and improvement. Second, there is already a number of applications like triggers or fast event filters, where high speed is more valuable than high purity. It is especially important for the experiments with high data rate, where it is simply not possible or too expensive to store all data for further accurate processing using classical algorithms.

\begin{figure}[!ht]
  \begin{center}
    \smallskip
    \includegraphics[scale=0.4]{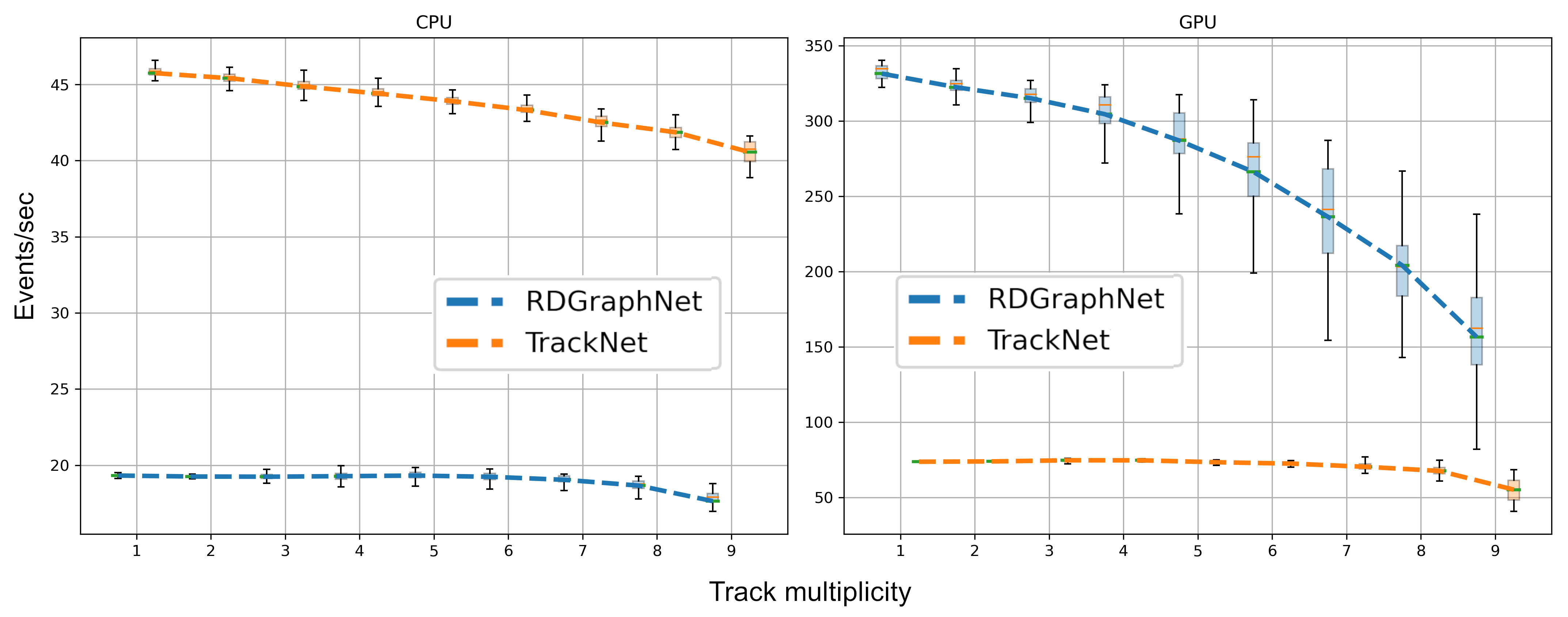}
    \smallskip
    \caption{
    Event processing speed comparison for BESIII between RDGraphNet and TrackNETv3, Python, Single-Threaded. 
    }
    \label{fig_proc_speed}
  \end{center}
\end{figure}

Finally, we measured the processing speed for RDGraphNet and TrackNETv3 models. These models have been run in similar conditions at the HybriLIT cluster (Intel(R) Xeon(R) Gold 6148 CPU @ 2.40GHz). The results are shown in  Fig. \ref{fig_proc_speed} and summarized in Table~\ref{tab_speed}. We see that the RDGraphNet model performs better than the local method with increasing event multiplicity, as expected. Surprisingly, the dependence on track multiplicity is drastically different, if models run on CPUs or GPUs.

%
%

\begin{table}[!ht]
  \begin{center}
    \caption{%
      Processing speed (number of events processed per second) for two models.
    }
    \label{tab_speed}
    \begin{tabular}{|p{0.5\textwidth}|c|c|}
      \hline
\ & RDGraphNet  & TrackNETv3 \\ \hline
Preprocessing (CPU, Python, single-threaded) & 19.28 & 43.91 \\ \hline
Preprocessing (CPU, C++, multithreaded) & 570 & --- \\ \hline
Inference (GPU) & 283.70 & 74.17 \\ \hline
    \end{tabular}
  \end{center}
\end{table}

\section{Conclusion}

\begin{enumerate}
\item
Two neural network algorithms based on deep learning architectures for track recognition in pixel and strip-based particle detectors are proposed. These are TrackNETv3 for local (track by track) and RDGraphNet for global (all tracks in an event) tracking. These algorithms were tested using the GEM tracker of the BM@N experiment at JINR (Dubna) and the cylindrical GEM inner tracker of the BESIII experiment at IHEP CAS (Beijing) and demonstrated  encouraging results, namely, >95\% recall and precision >74\% precision. These algorithms are general, and can be applied to any experiments using strip and pixel tracking detectors, including future experiments at the CEPC collider~\cite{CEPC}, the SPD experiment at the NICA collider~\cite{SPD}, and others that are running or planned.

\item 
We revised TrackNETv3 training approach and optimized the inference procedure using the Faiss library for efficient similarity search to reduce its algorithmic complexity. When running in the presence of a large number of fake hits, the network can be supplemented by a special classifier for filtering out false tracks, to increase the precision substantially. Our preliminary studies show that the TrackNETv3 network is capable to find tracks in events with multiplicity of up to 100 tracks.

\item
The processing speed of neural network algorithms looks promising and deserves further optimization. While the achieved purity of the `deep learning' tracking is still worse than the one of classical methods, there are already several applications like triggers or fast event filters, where high speed and high efficiency is more valuable than high purity. If necessary, classical methods like the Kalman filter can be applied afterwards to refine the search.

\end{enumerate}

\section{Acknowledgement}

The study was carried out with the financial support of the Russian Science Foundation No. 22-12-00109, https://rscf.ru/project/22-12-00109. 

Supported by the National Natural Science Foundation of China under Contract No. 11911530090.


\end{document}